# Polysaccharide/Surfactant complexes at the air-water interface – Effect of the charge density on interfacial and foaming behaviors

*Ropers M.H[1].[*], Novales B.[1], Boué F.[2] and Axelos M.A.V[1].*

[1] UR1268 Biopolymères Interactions Assemblages, INRA, F-44300 Nantes

[2] Laboratoire Léon Brillouin, UMR 12 CNRS/CEA-IRAMIS, 91191 Gif-sur-Yvette Cedex, France.

AUTHOR EMAIL ADDRESS: marie-helene.ropers@nantes.inra.fr

CORRESPONDING AUTHOR FOOTNOTE: Marie-Hélène Ropers , INRA, BP 71627, F-44316 Nantes Cedex 3, France. phone 0033 2 40 67 51 89, fax 0033 2 40 67 50 84

# ABSTRACT

The binding of a cationic surfactant (hexadecyltrimethylammonium bromide, CTAB) to a negatively charged natural polysaccharide (pectin) at air-solution interfaces, was investigated on single interfaces and in foams, versus the linear charge densities of the polysaccharide. Beside classical methods to investigate polymer/surfactant systems, we applied, for the first time concerning these systems, the analogy between the small angle neutron scattering by foams and the neutron reflectivity of films to measure in situ film thicknesses of foams. CTAB/pectin foam films are much thicker than that of the pure surfactant foam film but similar for highly and lowly charged pectin/CTAB systems despite the difference in structure of complexes at interfaces. The improvement of the foam properties of CTAB bound to pectin is shown to be directly related to the formation of pectin-CTAB complexes at the air-water interface. However, in opposition to surface activity, there is no specific behavior for the highly charged pectin: foam properties depend mainly upon the bulk charge concentration, while the interfacial behavior is mainly governed by the charge density of pectin. For the highly charged pectin, specific cooperative effects between neighboring charged sites along the chain are thought to be involved in the higher surface activity of pectin/CTAB complexes. A more general behavior can be obtained at lower charge density either by using a lowly charged pectin or by neutralizing the highly charged pectin in decreasing pH. .

1. Introduction

Polymer/surfactant assembly has received a constant interest over several decades since polymers may modify the properties of surfactant solutions[1,2,3,4]. Especially oppositely charged polymer/surfactant systems were widely studied as they strongly bind to each other and offer drastic changes in the properties of surfactant in the bulk and at interfaces[1]. One class of surfactants, quaternary ammonium salts, has been particularly considered in the polymer/surfactant studies to analyse the changes versus the effect of the alkyl chain length[5-13], the polymer structure which includes chain flexibility, charge density and hydrophobicity[13-23], the role of multivalent cations[24] and the ionic strength[25-28]. Among the negatively charged polymers that bind to this class of surfactants, pectin is a natural polysaccharide extracted from citrus and apple fruits at the industrial scale. Its structure is mainly constituted of galacturonic acid residues with some neutral sugars. It is schematically represented as a single molecule with long and smooth regions interspersed with short hairy zones (Figure 1). Around 100 (1→4)-α-linked-D-galacturonic acid residues constitute the smooth linear regions and a backbone of alternating galacturonic acid and rhamnose residues with frequent substitution of arabinose and galactose constitute the hairy regions[29, 30]. The percentage of galacturonic residues (COOH) esterified with methanol ($COOCH_3$), defines the degree of methylation (DM = [COOH]/([$COOCH_3$+COOH])) and the degree of charge ((100 − DM)/100). Pectin is widely used as gelling agent because of its interactions with cations especially with calcium for food applications[31]. The interest in such natural polymers is of importance in terms of stability of dispersed systems[32]. Pectin, as well as pectate, the fully de-esterified form of pectin, is also able to bind to cationic surfactants through its carboxylic groups[11, 12, 33, 34] but no binding is noticed at a degree of charge as low as 10%[35]. Coming now to foam properties, while some polysaccharides such as xanthan alter the foaming behaviour[18], some other polysaccharides enhance the stability of foams[36, 37] similarly to synthetic polymers[2,38]. Those effects may originate first from the gelling and thickening effects that increase the viscosity of the solution and decrease drainage velocity and gas permeability of the thin films[2,39], second from the complexation of polysaccharides with amphiphilic species at the interfaces that increase the surface viscosity[2,39] and third from the intrinsic properties of the polymer like the rigidity of the polymer backbone[2,18]. From a more general point view, as well documented in the case of synthetic polymers, the factors controlling the foam film stability are different in two different foam stages. In the foam formation stage, the surface rheological parameters as well as the density of the adsorbed layer play an essential role[2,39]. In the foam ageing stage, surfactant adsorption, surface tension, surfactant chain length and packing, capillary pressures in the Plateau border as well as stratifications properties of polyelectrolyte chains in the film bulk are claimed to govern the stability of the foam[40-43]. But more recently, the inhibition of the foam drainage has also been related to the rheology of the interface through the identification of a gel-like layer of polymer-surfactant aggregates[44]. Thus, the role of polysaccharides and more generally of polysaccharide/surfactant aggregates on the foam stability must be clarified. In this paper, we study the interfacial properties at the air-water interface of complexes formed between a quaternary ammonium salt and semi flexible negatively charged polysaccharides (pectin) with different charge densities and compare the foam properties of these complexes.

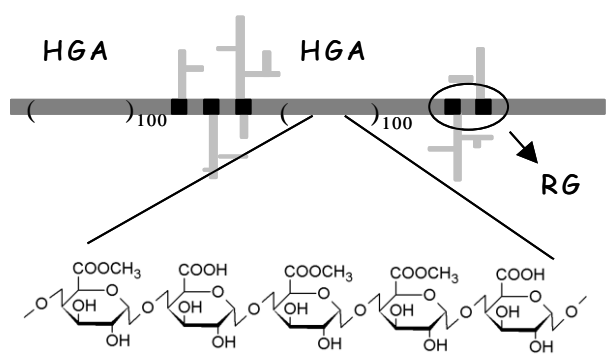

Figure 1. Schematic structure of the pectin backbone with two main polysaccharide domains homogalacturonan (HGA, dark grey bar) and rhamnogalacturonan (RG, black and dark grey zones) with branched chains of neutral residues of arabinose and galactose (light grey).

2. Materials and methods

2.1. Materials

Pectins provided by Copenhagen Pectin (Denmark) under the names Genupectin X-6011, X-6014 and X-6012 carry degrees of methylation of 37%, 52% and 77% corresponding to the linear charge densities of 1.01, 0.77 and 0.37 respectively (calculated according to literature data[31]). In the following, we will refer to the names P37, P52 and P77. These samples of pectin derive from the same (citrus) mother pectin through a basic de-esterification so that the charge is statistically distributed and they possess an identical backbone. They were purified to extract traces of divalent cations by following a procedure described by Dronnet et al.[45]. The free galacturonic acid content, evaluated from titration methods, amounts to 2.40, 1.88 and 0.88 mM per gram of dried pectin for P37, P52 and P77 respectively. Pectin powders were dissolved in the appropriate solution (ionic strength 25mM, water purified from a Millipore Milli-Q water system) at pH 5 under magnetic stirring overnight at 4°C. After solubilization, the pH was adjusted to 6.5 (except for other conditions specified in the text). The intrinsic viscosity of these pectin samples were found to be around 0.3 L/g. In our conditions of concentration, well below the semi-dilute regime, this value corresponds to viscosities of the pectin solutions close to that of water. Cetyltrimethylammonium bromide (CTAB) is a quaternary amine salt provided from Fluka and used without further purification.

It is fully soluble at 30 °C and has a critical micellar concentration of 0.18 mM as determined by isothermal titration calorimetry in 25mM of NaCl.

2.2. Solubility diagrams

The solubility diagrams of pectin-CTAB systems were established by varying the surfactant concentration between $10^{-7}$ and $10^{-3}$ M at a fixed galacturonic concentration (0.24mM) for each pectin in the following conditions: neutral pH (6.5), ionic strength (25 mM NaCl) and temperature 30°C. The weight concentrations of pectin that correspond to the galacturonic concentration of 0.24 mM are 0.1 g/L for P37, 0.13 g/L for P52 and 0.27 g/L for P77. Pectin solutions and CTAB solutions were mixed in equal volumes and stored at 30 °C for 7 days until phase identification. Phases were identified by visual inspection. We distinguished monophasic and biphasic phases and among the last ones, between turbid and precipitated phases.

2.3. Surface tension measurements

The interfacial properties of CTAB/pectin solutions were analyzed by measuring the surface tension of CTAB/pectin solutions at the air/water interface at 30 °C on a rising drop tensiometer (IT Concept, Longessaigne, France). Solutions were prepared in the same manner as for the phase diagram determination. The surface tension was measured through the shape analysis of an air bubble (6 μL) formed at the tip of a stainless steel curved needle dipped in the solution[46]. The complete adsorption of the surface active species and their rearrangement at the interface took at least 2 hours. Each point of the curve corresponds to the surface tension obtained after 2.5 hours, yielding a very weak residual variation ($d\pi/dt < 10^{-2}$ mN/min). The surface viscoelastic properties were determined by assessing the response of the interfacial tension to a sinusoidal variation of the bubble area (characterized by the pulsation ω and a constant variation of 1 mm$^2$ corresponding to the percentage of deformation ~7%). These experiments were performed after the adsorption of all surface actives species i.e. when the surface tension had reached its equilibrium value. In such experiments, the viscoelastic modulus ε is a complex number, with the real part that corresponds to the elastic contribution accounting for the recoverable energy stored in the interface and the imaginary part that reflects the loss of energy through any relaxation process occurring at or near the interface.

2.4 Foaming properties

Foaming experiments were conducted on a "Foamscan" apparatus developed by IT Concept (Longessaigne, FRANCE). The formation of the foam, the stability and the drainage of liquid from the foam were followed by conductivity and optical measurements. After calibration of the conductivity electrodes, 8 mL of solution in the sample cell was sparged with gaseous $N_2$ through a porous glass filter at a flow rate of 25 mL/min. The gas flow was stopped after reaching a final foam volume of 35 mL. The foamability corresponds to the time needed to reach 35 mL of foam volume. After stopping the bubbling, the free drainage of the foam was followed for 1200 s. Several parameters were automatically recorded by the "Foamscan" analysing software. The drainage of liquid from the foam was followed via conductivity measurements at different heights of the foam column. A pair of electrodes at the bottom of the column was used for measuring the quantity of liquid that was not in the foam, while the volume of liquid in the foam was measured by conductimetry in three pairs of electrodes located along the glass column. The volume of foam was determined by use of a CCD camera (Sony Exwave HAD). The liquid stability (LS) is the time needed to decrease by 50% the initial liquid volume in the foam. This parameter may be taken to characterise the drainage providing the starting volumes of foams are identical.

2.5 SANS neutron scattering on foam samples

Solutions of CTAB-pectin complexes were prepared at a concentration of $10^{-3}$ M in surfactant and at a galacturonic concentration of 0.24 mM, i.e. 0.1 g/L for P37 and 0.27 g/L for P77. The solvent was heavy water $D_2O$ (Eurisotop, France) in order to obtain a maximal contrast, defined as $\Delta Nb = Nb(i)-Nb(j)$ where Nb is the scattering length density and i and j correspond to two different compounds (Figure 2). The low concentration of pectin (0.27 g/L at the most) does not modify significantly the scattering length density of the solution in $D_2O$ compared to the one of $D_2O$, so that the latter can be considered as the scattering length density of the aqueous part comprised between the two interfacial films in foams. The intensity varying as the squares of the contrast ($\Delta Nb^2$), the interface CTAB/$D_2O$ interface is the major interface contributing to the intensity.

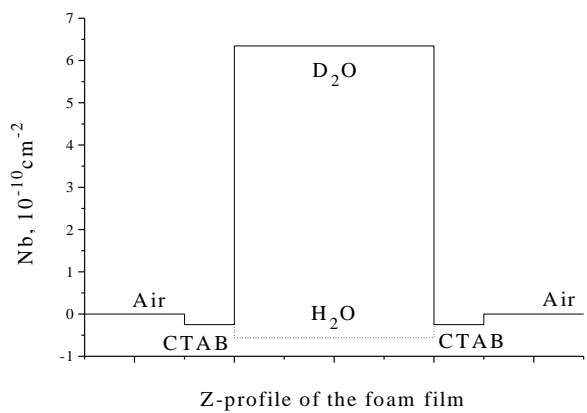

Figure 2. Profile of the scattering length density across a foam film of CTAB versus the solvent ($H_2O$ or $D_2O$). Scattering length density of CTAB was taken from the work of Taylor[6].

SANS foam measurements were performed at the Saclay Orphée reactor on the PACE spectrometer with detector cells forming circular rings. Two different configurations were used to cover a wide range of scattering vector modulii q, defined by $q = (4\pi \sin\theta)/\lambda$, where $2\theta$ is the scattering angle and $\lambda$ the wavelength. In the high-q configuration (0.0277 < q < 0.285 Å$^{-1}$), collimating was achieved with a first diaphragm of 22 mm located 2.5 m before the sample and a second diaphragm of 7 mm located just before the sample. The wavelength was 7 Å and the distance between the sample and the detector was 1 m. In the low-q configuration (0.004 Å$^{-1}$ < q < 0.0435 Å$^{-1}$), collimating was achieved with a first diaphragm of 12 mm situated 5 m before the sample and a second diaphragm of 7 mm just before the sample (here the foam column, see below). The wavelength was 10 Å and the sample-detector distance was of 4.5 m.

The foams were prepared in a foam cell, or column, very similar to that described in Axelos et al.[47]. The lower part of the cell is constituted of a quartz tube, which can be diametrally crossed by the neutron beam on a wide range of height. It is similar to the glass cylinder of the Foamscan apparatus (paragraph 2.4), with a larger diameter in order to increase the intensity arising from a higher number of foam films met by the beam (30 mm wide against 19 mm in Foamscan). The bubbling was adjusted to get a foam height of 10 cm. As soon as it was stopped, the neutron scattering intensity was recorded on the free-draining foam every 600 s for CTAB foams and 230 s for CTAB/pectin foams. In our conditions, the estimated number of films crossed by the beam is around 25 and even below 10 for some very dry foams. In parallel with the measurements on foams, a fraction of each sample solution was poured into a Hellma cell in order to measure the intensity scattered by the bulk solution.

Standard corrections for sample volume, neutron beam transmission, and incoherent scattering due to protons or deuterons were applied. The scattering of the empty cylinder was subtracted from the scattering of the cylinder filled with the foam and this difference was divided by the scattering intensity of 1 mm of water. Data configuration were corrected from the solvent contribution by subtracting a constant which takes into account the $D_2O$ contribution and the incoherent scattering due to the protons of CTAB and Pectin. We used the low q transmission data measured both in the foam and in the corresponding 2 mm CTAB or CTAB/Pectin solutions to estimate the total thickness of the solution in the foam crossed by the beam[47]. For dry foams this thickness was found to be around 1.5 mm which corresponds to a liquid fraction of about 0.05.

2.6 SANS data analysis

The analogy of SANS from a foam with multiwall reflectivity, and usual reflectivity by a thin layer, has been formerly evoked[47-49]. This new concept considers that a foam submitted to an incident radiation beam is a collection of M mirrors reflecting part of this radiation. Each mirror is one of the M foam cell walls. Here it is a neutron parallel beam (with an angular collimation $\delta\theta$), which hits each mirror indexed by the integer m with an incidence angle $\theta_m/2$, and is reflected at a "specular" angle $\delta\theta/2$, so that the total deflection is a scattering angle $\theta = \theta_m$. This is sketched in Figure 3.

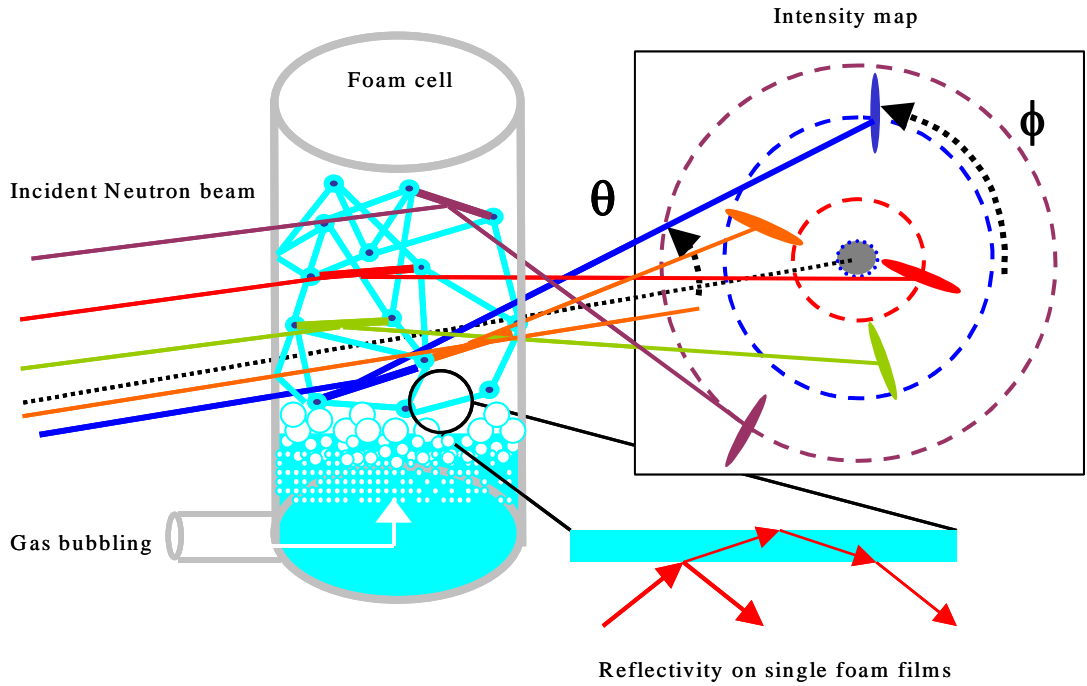

Figure 3. Sketch of a foam produced by bubbling in a cell with incident and reflected beams on some of the cell walls. θ is the angle between incident and reflected beam; ϕ is the azimuthal angle (both shown here for the blue ray).

Several walls could give a reflection at the same angle θ, at different angles φ so that the reflection spots are located on the same circle on the detector. Adding all contributions on all cells of the multidetector corresponding to the same deflection angle θ, corresponds to a radial average along this circle. This gives the scattered intensity for the corresponding scattering vector of modulus $q = (4\pi/\lambda)\sin(\theta_m/2)$, for the different values $\theta_m/2$. The result is equivalent to what obtained with a single mirror oriented at several angles $\theta_m$. In practice, the observed distribution of the $\theta_m$ values is continuous. This is due to the fact that each reflection spot is actually spread into a "spike" of large angular width $\Delta\theta_m$. This has been attributed to the fact that foam cell walls are slightly bent[47] as seen also on single films[49]. If the effective incident angle distribution is even wide and flat enough (this has to be kept in mind), the result is close to a reflectivity curve obtained by rotating continuously a single mirror in front of the beam.

The question we address here is: what is this curve for such mirror, i.e. a cell wall, which can be considered here as mostly constituted by a thin layer of water, with a thickness between 10 nm and 50 nm? It is relatively easy to calculate such curve using simple simulation software (available for example at www-llb.cea.fr). For a single dioptre, i.e. an infinite layer of species A with interface with vacuum, the rate of reflected intensity $R(q)$ shows, as a function of θ or better of q, two zones: 1) Between 0 and $q_C$, a "reflectivity plateau", for which $R(q) = 1$ ; the upper edge of the plateau, $q_C$ is the critical value; $q_c \sim (4\pi Nb)^{1/2}$ where $Nb$ is the scattering length density of the medium. This is a consequence of Descartes-Snell laws: below an angle $q_C$, there is no refracted beam, only a vanishing wave, so all incoming intensity is reflected. 2) For $q > q_C$, a fast decay reaching an asymptotic variation $q^{-4}$ (Fresnel law). Another consequence of Descartes-Snell laws: below an angle $q_C$, there is a refracted beam, so only a part of the beam is reflected.

When the layer thickness t becomes finite, fringes appear in the decaying part, with maxima separated by a period $\Delta q = 2n\pi/t$. This is not related to Descartes-Fresnel law, but to interferences existing between the fractions of the beam reflected at each film-vacuum interface. For large t values, this effect coexists with Descartes-Fresnel law, and the reflectivity plateau is not modified, i.e. $q_C$ is the same. When the thickness decreases, fringes are further apart from each other, and at the same time a new effect appears: the reflectivity plateau progressively vanishes. This is because the evanescent wave reaches the other interface, so a part of the beam is refracted. The first maximum tends asymptotically to $q = 2\pi/t$ when the thickness t tends to zero.

These effects can be smoothed by the angular width of the beam, some uncertainty on the thickness, etc. But they can be magnified graphically using a $q^4R(q)$ representation. In this case, when a a full reflectivity plateau exists for R(q) at low q, $q^4R(q)$ increases like $q^4$ in this region. When q reaches $q_C$, R(q) starts decreasing, and this results in a first maximum in $q^4R(q)$, observed at $q_C$. This maximum does not correspond directly to some interference fringes. At larger q, $q^4R(q)$ reaches a second maximum which corresponds to an interference fringe ; the following maxima at larger q have the same origin, i.e. fringes, and should be separated by $\Delta q = 2\pi/t$. We show in Figure 4 a series of reflectivity curves at decreasing thickness, for the same Nb ($D_2O$) and in Figure 5 the position of the first maximum versus the thickness of the film between 50 and 500 Å. Both Figures indicate that at large thicknesses (over 300Å), the position of the first maximum of $q^4R(q)$ does not depend on the thickness and is very close to $q_C$ while at low thicknesses (t < 300Å), it disappears.

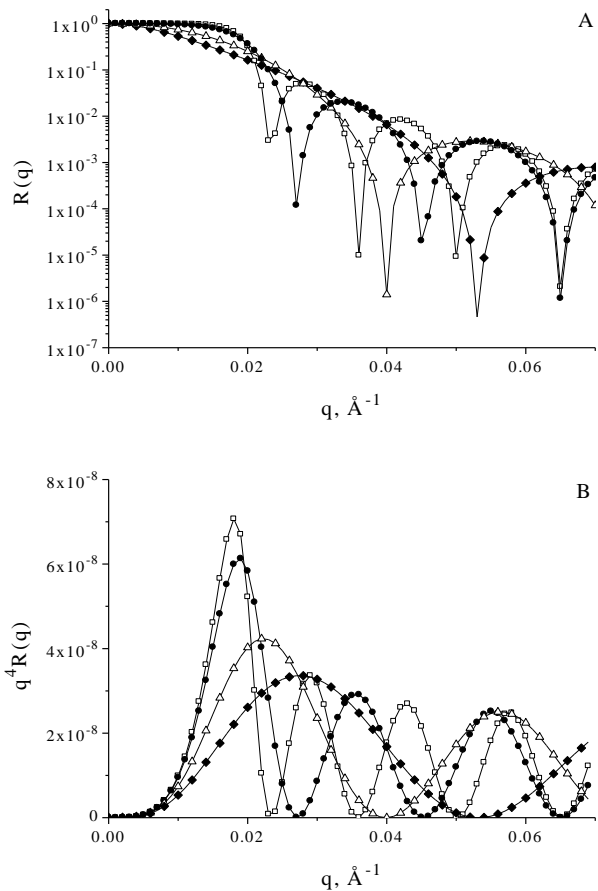

Figure 4. Reflectivity curves of foam films of thickness (□) 400 Å, (●) 350 Å, (△) 175 Å and (◆) 125 Å represented as R(q) (Figure A) and $q^4R$ versus q (Figure B).

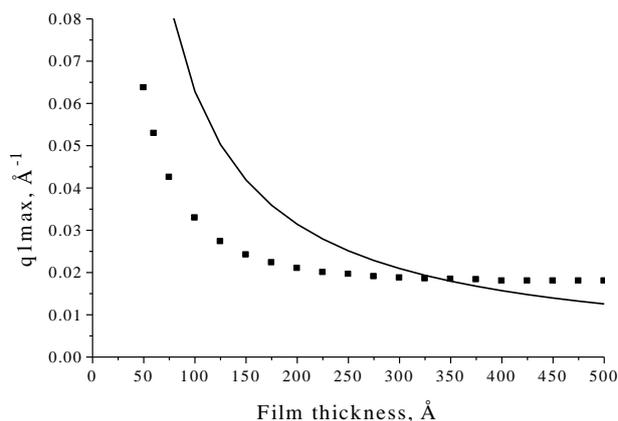

Figure 5. Values of the scattering wavevector q of the first maximum of the reflectivity curves (shown as $q^4R(q)$) as a function of the thickness of a single $D_2O$ film. The solid line refers to the function $2\pi/t$ where t is the thickness of the film.

We have used in this paper the analogy between SANS by a foam with multiwall reflectivity and reflectivity by a thin layer to analyse the SANS measurements on CTAB and CTAB/Pectin foams, and to deduce the thickness of the films.

3. Results and Discussion

3.1. Solubility diagrams of CTAB/pectin assembly

The observations are summarised in Figure 6. At very dilute surfactant concentrations (CTAB < $10^{-4}$ M), all mixtures are clear and transparent. Pectin-CTAB complexes are completely soluble. As the surfactant concentration is increased, the phase behaviour differs according to the degree of charge of pectin. In the P77/CTAB system, the solutions remain clear and transparent on the whole range of concentration investigated. For the solutions containing pectins P52 and P37, the solutions become turbid in a reduced concentration range close to the charge neutralization point. This is usually attributed to the decrease in solubility of the complex. For P52, as the concentration in surfactant is further increased, the solutions become clear again. The resolubilisation of the complex is usually ascribed to the excess of charge that accumulates on the complex. In the case of P37, the onset of precipitation occurs just above the neutralization point and there is no resolubilisation.

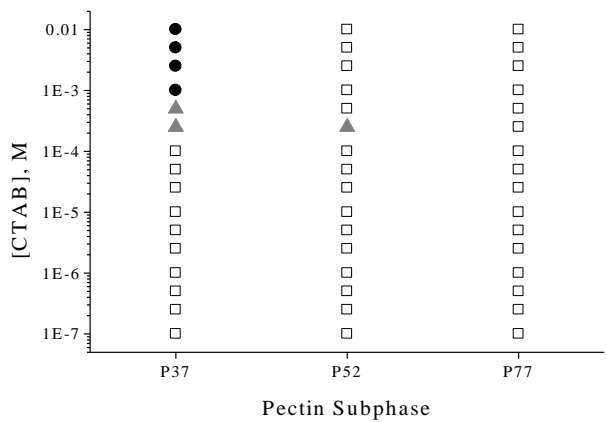

Figure 6. Solubility diagrams of pectin/CTAB complexes at 30°C. Empty squares represent clear and transparent solutions, grey triangles correspond to turbid solutions and the plain circles to precipitates. The pectin concentrations are 0.1, 0.13 and 0.27 g/L for P37, P52 and P77 respectively. The total pectin charge concentration (0.24 mM in all cases) is negligible in comparison with the charge concentration given by the salt (NaCl 25 mM).

3.3. Interfacial properties of CTAB/pectin complexes

3.3.1 Effect of DM

The formation of surface active macromolecular species was evidenced by measuring the surface tension during spontaneous adsorption of pre-formed surfactant/pectin complexes. The curves displaying the surface tension at equilibrium as a function of surfactant concentration are shown in Figure 7.

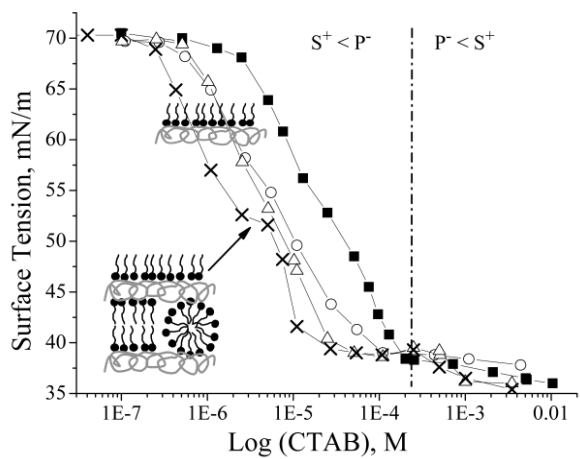

Figure 7. Surface tension curves of CTAB (■) and CTAB/pectin solutions at 0.24 mM in COO$^-$ for (✕) P37 0.1 g/l, (△) P52 0.13 g/L and (○) P77 0.27 g/l. The dashed line corresponds to the equivalent point of charge separating the domains where the surfactant molecules (S$^+$) or the pectin molecules (P$^-$) are the major part. The arrangement of the polymer within the surfactant layers is unknown and the drawings are only indicative on the adsorption behaviour (monolayer- or multilayer-like).

The surface tension values of CTAB solutions start to decrease at $2.5 \times 10^{-6}$ M and the micellar state is reached at $1.8 \times 10^{-4}$ M, in accordance with isothermal titration calorimetry measurements. Over these two decades, the surface tension decreases monotonously. The presence of pectin has a drastic effect. Several key points must be noted.

(i) a more efficient lowering of surface tension is observed in the order P37>P52>P77.

(ii) CTAB/P37 solutions behave distinctly from CTAB/P52 and CTAB/P77: it shows a non monotonous decrease of the surface tension with concentration, with a short plateau (here at 52 mN/m) similarly to the systems dodecylammonium chloride/τ-carrageenan[15] and dodecylammonium chloride/polystyrenesulfonate[51]. By analogy with these previous results from the literature data, we propose that the plateau at 52 mN/m corresponds to the onset of micelle formation on the polymer (critical aggregation concentration, CAC). The fact that the plateau does not exist for P52 and P77 can be directly related to their lower charge density.

(iii) above the equivalent point of charge (0.24 mM, where the charge brought by surfactant, $S^+$ become larger than the one brought by pectin, $P^-$), the surface tension curves almost coincide. The excess monomers of surfactant molecules have come to interface and constitute mainly the interfacial film. The onset of this plateau at around 38mN/m occurs at increasing concentration in the order of DM, i.e., $1.5 \times 10^{-5}$ M for P37, $3.5 \times 10^{-5}$ M for P52 and $1 \times 10^{-4}$ M for P77 while it starts at $1.8 \times 10^{-4}$ M for pure CTAB.

Below the equivalent point of charge, between 70 and 52 mN/m, the surface tension curve of CTAB/P37 solutions is shifted to approximately tenfold smaller surfactant concentrations in comparison to the pure surfactant curve, while this factor is only 4 for CTAB/P52 and CTAB/P77 solutions. The slopes of the three curves are however similar in this tension range. The species located at the interface are thus presumably of the same type. According to the review of Taylor et al. about polymer/surfactant assembly at interfaces, two kinds of adsorption behaviour can be found in polymer/surfactant systems and reveal the types of aggregates adsorbed at the interface[3]. These convenient schematic diagrams however apply only for zero ionic strength. To take into account the addition of electrolyte and the fact that our polymers are not fully charged, we refer to the case of $C_{12}$TAB/NaPSS (dodecyltrimethylammonium bromide and sodium polystyrenesulfonate)[51]. In this case, the addition of electrolyte (0.1 M NaBr) enlarges the concentration range over which the tension varies. Moreover, the plateau associated to the onset of the micelle formation on the polymer is less pronounced[3, 51]. It is initially located at 45mN/m and goes up to around 51 mN/m.

The similarity of the curves for $C_{12}$TAB/NaPSS system in the presence of electrolyte and our CTAB/P37 system suggests that the aggregates are probably of the same type or result from the same type of mechanism. In the lower surfactant concentration range (at concentrations smaller than $2.5 \times 10^{-6}$ M, surface tension value larger than the plateau), the surface active species would be thus a monolayer of surfactant/polymer complexes (described as Type 2 adsorption behavior by Taylor[3]). At higher concentrations (between $5 \times 10^{-6}$ and $2.5 \times 10^{-5}$ M, surface tension value lower than the plateau), polymer/surfactant aggregates are described at the interface as a multilayer complex (scheme in Figure 7) with an adsorbed monolayer of surfactant and underneath a complex structure of surfactant molecules organised in micelles or elongated micelles adsorbed with polymer to the underside of the monolayer (named Type 1 adsorption behaviour by Taylor[3, 51]). In this range, the slope is steeper confirming the fact that the polymer/surfactant multilayer is more densely packed than the monolayer surfactant/polymer complexes.

As concerns CTAB/P52 and CTAB/P77, the lack of CAC and the similarity of the slopes until the free micellization in solution takes place, prove that the surface active aggregates are surfactant/polymer complexes arranged in a monolayer. The disappearance of the intermediate plateau at 52 mN/m with increasing DM has been also encountered for $C_{12}TAB$/PAMPS system (partially charged acrylamide polymer) and has been interpreted in terms of low cooperativity between alkyl chains[52]. Surfactant molecules bounded to the polymer are too far from each other to be assembled into stable micelles wrapped by polymer.

3.3.2 Effect of the linear charge density

To confirm this interpretation, we compare in Figure 8 different systems where negative charges of pectin are brought in solution either through weight concentration or by varying the pH. The titration curve of P37 tells us that an equal $COO^-$ molarity is brought by: P37 at 0.1 g/L pH 4.2, P77 at 0.1 g/L pH 6.5, and P37 at 0.04 g/L pH 6.5. The first striking result is that P37 at pH 6.5 at 0.04 g/L displays the same 52 mN/m tension plateau than at 0.1 g/L (the width of the plateau is just reduced). On the contrary, the complexes formed with P37 at acidic pH (4.2, 0.1 g/L) do not display any plateau and exhibit a surface activity close to the P77 charge equivalent solution. So decreasing the charge by decreasing pH or by increasing the methylation appears equivalent, in agreement with the fact that both acid-base dissociation and methylation are statistical along the chain. We conclude that CTAB/P37 complexes at pH 4.2 form a monolayer at the air-water interface while at neutral pH they form either a monolayer or multilayers according to the concentration range.

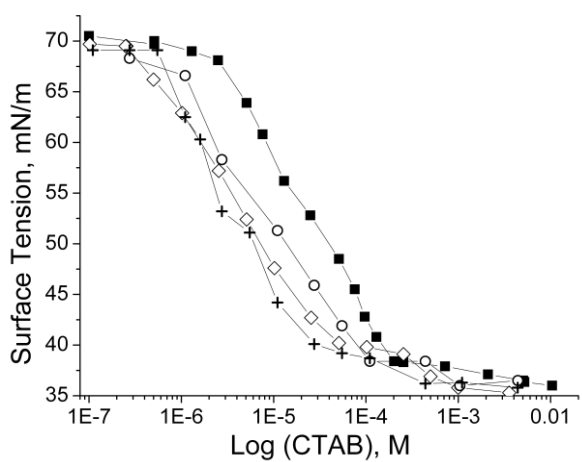

Figure 8. Surface tension curves of CTAB (■) and CTAB/pectin solutions with (+) P37 at 0.04 g/l and pH 6.5, (◇) P37 at 0.1 g/l pH4.2 and (○) P77 at 0.1 g/l pH 6.5.

The degree of methylation (DM) has therefore a direct role in the surface activity of pectin/surfactant complexes. The decrease of the concentration does not modify the behaviour. But if we modify the distribution of charge on pectin by (statistical) protonation, the surface-active behaviour changes. We assume that there exists a strong cooperativity between surfactant molecules fixed on long segments of fully charged galacturonic residues. The degree of charge of P37 (63%) enables the occurrence of enough long charged segments that impact on the binding and the resulting assembly. As a consequence, the distinction between highly methoxylated pectins (like P37) and lowly methoxylated pectins (like P52 and P77), which is usually used for calcium binding[31], applies here for the complexation with surfactants. The flexibility of the polymer is also a positive parameter that enables the micellization of bound surfactant neighbours by the wrapping of the polymer.

### 3.3.3 Dilatational rheological experiments

Figure 9 shows the elastic modulus at equivalent charge concentrations, for two surface tension values: 65 mN/m, from which we infer from the discussion above that the structure is of the monolayer-type for all systems, and 47 mN/m, for which the monolayer is replaced by a multilayer-type structure in the case of P37. It appears clearly on Figure 9 that pectin strongly contributes to change the viscoelasticity of CTAB interfacial film.

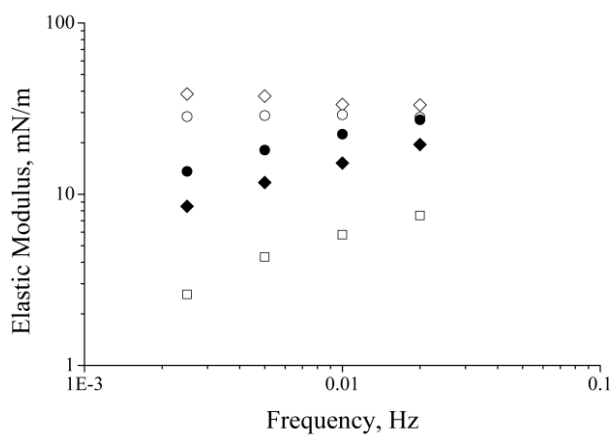

Figure 9. Elastic modulus (in mN/m) of CTAB-pectin complexes adsorbed at the air-water interface at equivalent bulk charge concentrations, at surface tension values of 65 mN/m (◇, ○, □) and 47 mN/m (◆, ●). Symbols are for (◇, ◆) CTAB-P37, (○, ●) CTAB-P77 and (□) CTAB.

At the surface tension 65 mN/m, both CTAB-pectin samples gives a flat response, indicative of an elastic behaviour ("solid-like"), at least in the investigated frequency range, while CTAB alone behaves like a liquid. Pectin is thus responsible for the connectivity along the monolayer. The higher value of the modulus found for CTAB-P37 than for CTAB-P77 agrees with the previous idea that a higher surface charge density between P37 and CTAB at the interface makes higher the cross-linking density, hence the connectivity. At the surface tension 47 mN/m, moduli are lower than those obtained at 65 mN/m and both CTAB-pectin samples exhibit a liquid-like behaviour, as indicated by the variation of the modulus curve with frequency. Those results suggest that the layer of CTAB-pectin (monolayer or multilayer-type) is less homogeneous. It may be regarded as patches with no connectivity between them. In the case of P37, the change in elasticity from solid to liquid-like behaviour may be related to the change in structure at interfaces, observed in Figure 7. The binding of finite objects to the monolayer of CTAB mediated by pectin molecules results in the lost of homogeneity and connectivity. In the peculiar case of P77, the change in elasticity from solid to liquid-like behaviour is attributed to defects in the accumulation of the lowly charged pectin molecules at the interface. Indeed, the orientation of charges carried by pectin toward the film of CTAB is limited by the semi-flexibility of the pectin chain and the bulkiness of all uncharged sites between two next nearest charged sites along the molecule. This induces the formation of small domains of CTAB monolayers next to CTAB/P77 domains at the interface (local phase separation).

3.4. Foaming properties of CTAB/pectin assembly

Let us first point out that pectin solutions do not foam: no foams were obtained for concentrations as high as 10 g/L, which is around hundred times higher than the concentrations investigated in this paper. As a consequence, all foaming properties will arise from CTAB or CTAB/pectin complexes, and we will compare the two systems. In the case of CTAB solutions ($5 \times 10^{-5}$, $10^{-4}$, $2.5 \times 10^{-4}$, $5 \times 10^{-4}$ and $10^{-3}$ M), the initial volume of foam (35 mL) was reached in less than 70 seconds except for the lowest concentration where this time is 87 seconds. At that concentration, the foam was not stable and disrupted in less than 5 minutes. The other foams were more stable with a final volume of 30 mL at the end of the experiment (Figure 10). The liquid stability (LS) increased from 53 seconds to 88 seconds as a function of CTAB concentration (Table 1). However, the general trend is that foams produced from pure CTAB solutions are not very stable and characterized by a high speed of drainage.

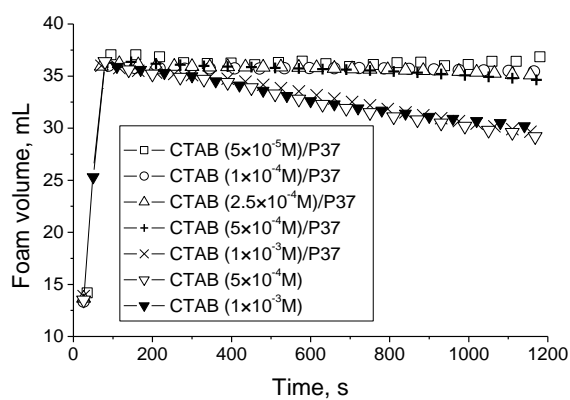

Figure 10. Evolution of the foam volume for CTAB solutions and CTAB/P37 (0.1 g/L pH 6.5) mixtures at various CTAB concentrations ranging from $5\times10^{-5}$ M to $10^{-3}$ M.

| Concentration | **CTAB** | **CTAB/P37** |
|---|---|---|
| **$5\times10^{-5}$ M** | 53 | 43 |
| **$10^{-4}$ M** | 60 | 127 |
| **$2.5\times10^{-4}$ M** | 81 | 193 |
| **$5\times10^{-4}$ M** | 88 | 247 |
| **$10^{-3}$ M** | 80 | 84 |

Table 1. Liquid stability (in seconds) for pure CTAB and CTAB/P37 mixtures as a function of CTAB concentration.

In the presence of pectin, the foaming properties of CTAB are strongly modified (Figure 10). Only pectins of DM 37 and 77% were used, since P52 acts on surface tension similarly to P77, as seen above. Let us first consider the case of P37 at pH 6.5 in the presence of increasing CTAB concentrations.

3.4.1 Effects at variable CTAB contents

For a CTAB concentration of $5\times10^{-5}$ M, the foamability is improved. While the foam from pure CTAB solution disappears in a few minutes, the foam formed from CTAB/P37 solutions at $5\times10^{-5}$ M does not completely disrupt. At the end of the experiment, the foam is still visible in the glass column; however it is constituted of quite larger bubbles. For larger CTAB concentrations, from $10^{-4}$ to $5\times10^{-4}$ M, the addition of P37 leads to the formation of very stable foams with the same bubble size as for pure CTAB solutions. Moreover, the foam volume remains constant (Figure 10) and the drainage is considerably slowed down (Table 1). In this concentration range, the liquid stability is increased by a factor 2 to 3 in the presence of P37. At larger CTAB concentrations ($\geq 10^{-3}$ M), the stability decreases sharply down, back to values comparable to pure CTAB ones: the foam volume (Figure 10) and liquid stability values (Table 1) are indeed similar in the absence and in the presence of pectin. Pectin has no effect on the foam properties. Note that at this concentration, the surfactant is in large excess compared to the galacturonic content ([COO$^-$] = 0.24 mM, [CTAB] = 1 mM, so $S^+/P^-$ = 4). One can thus reasonably assume that all CTAB/P37 aggregates have precipitated in this concentration range; so CTAB imposes its foaming behaviour.

3.4.2 Effect of pectin charge: Equivalence of pH and weight concentration.

As the best foaming properties of CTAB/P37 mixtures were obtained at a CTAB concentration of $5\times10^{-4}$ M, all other foams produced from CTAB/Pectin solutions will be mainly compared at this CTAB concentration. The concentration in pectin was adjusted in order to compare systems possessing the same amount of charge in bulk, similarly to the surface tension measurements. In practice, two ways were used to adjust P77 and P37 concentrations at the same bulk charge concentration: either by (i) decreasing the weight concentration of P37, or (ii) by decreasing pH for P37. To meet these two conditions, (i) CTAB/P37 mixtures at 0.1 g/L pH 6.5 and CTAB/P77 mixtures at 0.27 g/L pH 7 were compared on one hand, and (ii) CTAB/P37 mixtures at 0.1 g/L pH 4.2 and CTAB/P77 mixtures at 0.1 g/L pH 7 were compared on the other hand.

Figure 11 shows that whatever the way to match the concentration of charge, we observe that the volume of the foams and their liquid stability (Table 2) are very similar at equivalent bulk concentrations of charge. The liquid volume fraction within the foams, shown in Figure 12, also displays equivalence between the two ways of matching the concentration of charge. In the investigated concentration range $5\times10^{-5}$ < [CTAB] < $10^{-3}$ M, the percentage of liquid volume in the foam of both CTAB/P77 0.1 g/L and CTAB/P37 0.1 g/L evolves similarly. It increases with increasing CTAB concentration up to $5\times10^{-4}$ M, then decreases at higher concentrations. At the same time, we see a shift in y-axis: we can assume that it reflects the different concentrations of aggregates. Indeed, the decreasing bulk concentration of charge (from DM 37 to 77 both at 0.1 g/L) decreases the concentration of CTAB/pectin complexes, so it should decrease, by the way, their contribution to the foam stability and the water retention within the foam.

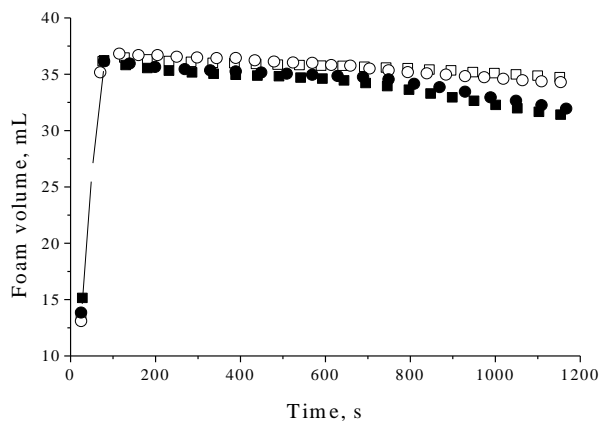

Figure 11. Evolution of the foam volume of CTAB/Pectin foams for (□) P37 0.1 g/L pH 6.5, (○) P77 0.27 g/L pH 6.5, (■) P37 0.1 g/L pH 4.2, (●) P77 0.1 g/l pH6.5 at a concentration in CTAB of $5\times10^{-4}$ M.

| pH and Pectin concentration | Time (s) |
| --- | --- |
| P37  0.1 g/L  pH 6.5 | 247 |
| P77  0.27 g/L  pH 6.5 | 223 |
| P37  0.1 g/L  pH 4.2 | 154 |
| P77  0.1 g/L  pH 6.5 | 183 |

Table 2. Liquid stability for CTAB/Pectin mixtures at a concentration in CTAB of $5\times10^{-4}$ M and in different pectin concentration and pH conditions.

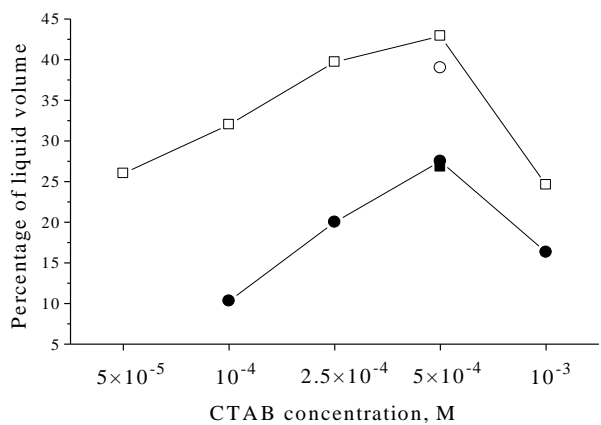

Figure 12. Percentage of liquid volume in the foam of CTAB/Pectin mixtures for (□) P37 0.1 g/L pH 6.5, (○) P77 0.27 g/L pH 6.5, (■) P37 0.1 g/L pH 4.2 and (●) P77 0.1 g/l pH6.5, observed 10 min after the end of foaming (compared to the liquid volume in the foam just after the end of bubbling).

It appears clearly that points obtained at equivalent bulk charge concentrations are equal or almost equal. The decrease in pH makes the foaming behaviour of CTAB/P37 solutions similar to that of CTAB/P77 solutions at pH 6.5. This behaviour is in accordance with the discussion in Section 3.3, *i. e.* the decrease in pH promotes the formation of CTAB/P37 complexes that resemble CTAB/P77 complexes (Figure 8). The foaming properties are also similar when considering the charge matching by varying the weight concentration. However, we can note a small difference, at same charge matching, between empty circles in Figure 12 (CTAB/P77 at 0.27 g/L) and empty squares (CTAB/P37 at lower weight fraction, 0.1 g/L ). The former gives foam slightly drier. We would expect that the liquid volume increases with the amount of polysaccharides due to the hydration around the polymer. However we observe the reverse trend. This can reflect the differences in CTAB/pectin aggregates structure and their different role in the Plateau borders. We have inferred that the structures of aggregates are different from the surface tension measurements (Figure 7). CTAB molecules stack to P77 as individual molecules, due to the far distance between charges but constitute less homogeneous layers, while CTAB may stack as grouped into micelles on P37. Micelles wrapped by pectin are surrounded by a higher rate of hydration and this may explain the higher rate of liquid volume in the foam formed with P37 at pH 6.5. This small difference is in any case minor in comparison to the enhancement of the foaming properties of CTAB solutions by pectin.

Finally, the liquid volume in the Plateau borders can also be in balance with the film thickness. We will compare in section 3.5 these thicknesses as measured in situ by neutron scattering.

3.4.3 Best foaming and large aggregates

The occurrence of best foaming properties at the phase boundary between soluble surfactant-polymer complexes and insoluble complexes has been reported formerly[36]. This phenomenon occurs for CTAB/P37. The best foaming properties are encountered at concentrations close to the insolubility of complexes (at twice the equivalence of charge *i.e.* [COO$^-$] = 0.24 mM and [CTAB] = 5×10$^{-4}$ M). Insolubility of complexes can be attributed to a very large size. The best foaming properties of CTAB/P77 mixtures were also encountered at the same concentration. However there is no formation of insoluble complexes in this CTAB/P77 case, hence no strong increase of aggregate sizes (unless they do not reach sizes large enough to produce neither precipitation nor turbidity). As a consequence, the improvement of the foaming properties of CTAB/Pectin by the formation of large aggregates is probably not the only parameter affecting foam properties. In addition to this, it is noteworthy that the concentration in CTAB leading to the best foaming properties of CTAB/P77 at 0.1 g/L is the same as that encountered for CTAB/P37 at 0.1 g/L (Figure 12) while the concentration of free micelles and CTAB/pectin complexes are different.

3.4.4 Chain rigidity effects: comparison of pectin with xanthan.

Although pectin and xanthan are polysaccharides, this study, compared to that of Langevin[2] on xanthan proves that pectin and xanthan act oppositely on the foaming properties of alkyltrimethylammonoium bromide. Pectin improves the stability of foam properties of CTAB, while xanthan alters the foamability of cationic surfactants. This can be interpreted in terms of rigidity[18]. The persistence length of xanthan is around 400 Å[53], approximately ten fold higher than the persistence length of pectin (50Å)[54].

3.5. Small Angle Neutron Scattering measurements

The neutron scattering curves obtained on CTAB foams were recorded as soon as the bubbling was stopped (Figure 13).

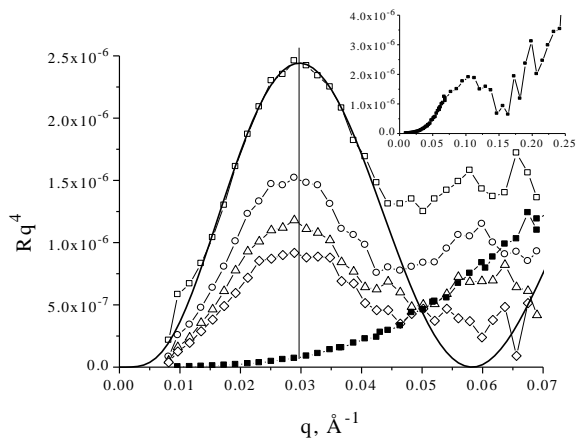

Figure 13. Scattering curves of CTAB foams represented as $Rq^4$ versus q. Curves were obtained after a drainage of respectively (□) 600 s, (○) 1200 s, (△) 1800 s and (◇) 2000 s from the top to the bottom. The straight line represents the reflectivity curve for an air/D$_2$O/air interface, 113 Å thick and possessing a scattering length density of $6\times10^{-10}$ cm$^{-2}$. The scattering curve of CTAB solutions is shown as filled squares in the same plot as the scattering curves of CTAB/foams and as an insert on a larger q-range to show the peak corresponding to micelles at round q = 0.12 Å$^{-1}$.

The curves $q^4R(q)$ versus q exhibit a maximum at around q = 0.03 Å$^{-1}$, the intensity of which decreases with increasing drainage. This is probably due to the decrease of the number of films (disproportionment or film rupture). The analogy between SANS by foams and reflectivity of single films developed in section 2.6 is used to fit the SANS curves of CTAB foams using reflectivity programs. The fitting of the curves was performed assuming a simple layer air/D$_2$O/air: we observed that when adding a CTAB layer, no sensible variation of Nb or thickness had any significant influence on the fits. The fitting was then performed in determining the value of the thickness that matches at the best the first maximum and the minimum of the reflectivity curve in the q$^4$R(q) representation.

For pure CTAB solutions foams, the freely-drained films are 113 ± 2 Å thick. Comparing with disjoining pressure measurements on salt-free CTAB solutions at the CMC[40], we note that a similar thickness (around 140 Å) is observed at high pressures, equivalent to the one necessary to get a dry foam film. This thickness corresponds to a Common Black Film and is far higher than the thickness of two CTAB monolayers, which would be closer to a Newton Black Film (2×21 Å according the work of Lu et al.[55]). We can assume that our foam wall is composed of two monomolecular films of CTAB separated by several layers of micelles or bilayers (C ≥ CMC). The film does not thin over the time scale of 45 minutes, since no additional maximum in q$^4$R(q) is observed at higher q-values. The increase in intensity which we can see above 0.05 Å$^{-1}$ (see Insert of Figure 13 for a larger q range) is the same as observed from scattering of CTAB micelles in solution, so it can be attributed to the micelle contribution from the aqueous phase in the Plateau borders.

When pectin binds to CTAB, the first maximum of the SANS curve of the foams is seen at lower q-values (around 0.018 Å$^{-1}$, Figure 14). This q-value does not change after more than one hour of drainage and second maximum appears, signalling a thicker foam film than for CTAB solutions foams. The curve of CTAB/P37 (not shown for clarity) exhibits exactly the same shape than that of CTAB/P77. Using again the SANS-reflectivity analogy for foams, the fit to a reflectivity curve of a single film gives a thickness of 300 ± 10 Å. In this case, the two interfacial monolayers are separated by free molecules and micelles wrapped by pectin. The amount of these has been estimated in determining the CMC of CTAB by conductivity measurements in the presence of pectin (without salt). It indicates that the formation of free micelles should occur at $3.5 \times 10^{-4}$ M in presence of P77. At $10^{-3}$ M, the film bulk should be constituted of free surfactant molecules for $1 \times 10^{-4}$ M and surfactant molecules localised in micelles wrapped by the polymer for roughly $2.5 \times 10^{-4}$ M. The similarity of the thickness of the foam film in CTAB/P77 and CTAB/P37 at equivalent charge concentration can be directly related with their similar foaming behaviour (Figure 11). It is noteworthy that the thickness of 300 Å is similar to the ones measured by film disjoining pressures experiments[40, 56] in several polysaccharide or synthetic polymers/surfactant systems. This similarity applies when films are in the drier state, i.e., when the pressure applied on the foam films is high.

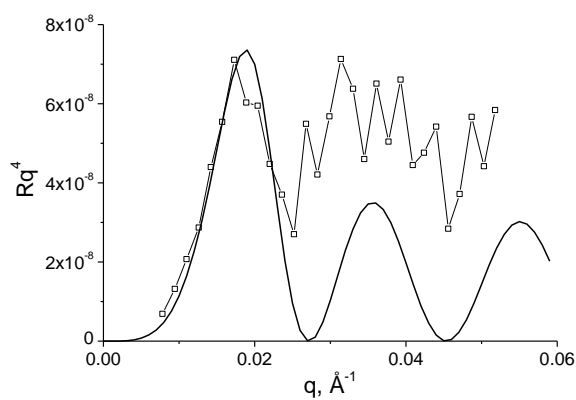

Figure 14. Scattering curve (□) of CTAB/P77 foams represented as $Rq^4$ versus q and recorded after a free drainage of 3900 s. The straight line (—) represents the reflectivity curve for a foam film 300 Å thick and possessing a scattering length density of $6 \times 10^{-10}$ cm$^{-2}$.

4. Conclusions

Binding of pectin to trialkylammonium surfactants is considered at the level of interfacial and foaming properties, in agreement with solution properties which are already available in the literature, in particular limiting DM for complexation[35], binding degree[11, 33] and rheological properties[34]. We have shown that pectin acts on the interfacial behaviour of alkyltrimethylammonium surfactant similarly to several other negatively charged polysaccharides. The complex formed in bulk between the negatively charged pectin and the cationic surfactant CTAB at a constant bulk charge concentration has a surface activity evolving with the degree of charge. The lower the DM, the stronger the complex surface activity (CTAB/P37 > CTAB/P52, CTAB/P77). Moreover, the modulation of the bulk charge by pH alters the surface activity of the low DM pectin/CTAB complexes. All results show that this behavior exhibits strong similarities with the affinity of pectin against calcium. In both cases, the distribution of charges along the pectin chain has a role in the cooperative phenomena.

For high DM pectins (DM 52% and 77%), the interface is thought to be covered by a thin layer of surfactant with the polymer anchored underneath, whatever the surfactant concentration. At low DM (37%), the same thin monolayer of pectin/CTAB is found at low concentrations and a more compact organisation of CTAB and pectin P37 is encountered at higher concentrations from the onset of micellisation on the biopolymer. In all cases, when the complexes are of the monolayer type, CTAB-P37 and CTAB-P77 films at the interface behave like a gel.

Unlike the much more rigid polysaccharide xanthan, pectin improves significantly the stability of foam properties of CTAB. In contrast to surface activity, no significant effect of DM on foaming properties is observed as far as the bulk charge concentration is matched. This is confirmed by measurements of the thickness by SANS. The foams are constituted by films of the same thickness that enlarges from 110 to around 300 Å in presence of pectin. The description of the interfacial layer of single interfaces is not a perfect mirror of the foam properties. The description of *in situ* interfaces, even difficult to achieve, must be pursued. In this sense, small angle neutron scattering proves here its interest to describe the film in foams.

ACKNOWLEDGMENT

Copenhagen Pectin is acknowledged for providing GenuPectin batches of various degrees of charges. We thank Dominique Guibert for establishing the phase diagram of CTAB/pectin solutions.

Table of Contents Graphic

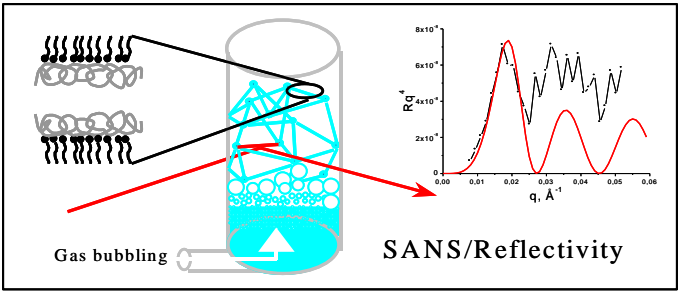